# Revision 1

# Humans plan for the near future to walk economically on uneven terrain

Osman Darici[1,*], Arthur D. Kuo[1,2]

[1]Faculty of Kinesiology and [2]Biomedical Engineering Program, University of Calgary, Calgary, Alberta, Canada

**\*Corresponding email:** osman.darici1@ucalgary.ca

**This PDF file includes:**

> Main Text
> Figures 1 to 8

## Abstract

Humans experience small fluctuations in their gait when walking on uneven terrain. The fluctuations deviate from the steady, energy-minimizing pattern for level walking, and have no obvious organization. But humans often look ahead when they walk, and could potentially plan anticipatory fluctuations for the terrain. Such planning is only sensible if it serves some an objective purpose, such as maintaining constant speed or reducing energy expenditure, that is also attainable within finite planning capacity. Here we show that humans do plan and perform optimal control strategies on uneven terrain. Rather than maintain constant speed, they make purposeful, anticipatory speed adjustments that are consistent with minimizing energy expenditure. A simple optimal control model predicts economical speed fluctuations that agree well with experiments with humans (N = 12) walking on seven different terrain profiles (correlated with model $\rho = 0.55 \pm 0.11$, $P < 0.05$ all terrains). Participants made repeatable speed fluctuations starting about six to eight steps ahead of each terrain feature (up to ±7.5 cm height difference each step, up to 16 consecutive features). Nearer features matter more, because energy is dissipated with each succeeding step's collision with ground, preventing momentum from persisting indefinitely. A finite horizon of continuous look-ahead and motor working space thus suffice to practically optimize for any length of terrain. Humans reason about walking in the near future to plan complex optimal control sequences.

## Significance Statement

Humans obviously look ahead as they walk, but it is unknown what adjustments they make for uneven terrain and for what objective. We show that humans purposefully (albeit subconsciously) plan and adjust their forward speed ahead of upcoming terrain, as predicted by minimization of energy expenditure. This research shows how the dynamics, energetics, and control of walking are predictable. It also reveals that humans have greater capacity to anticipate and reason about the body's interactions with the world than previously understood. They use this capacity to plan complex control sequences that save them energy.

## Introduction

Walking over irregular terrain presents considerable control challenges compared to level ground (Figure 1A). Whereas level walking is periodic and largely explained by minimizing energy expenditure (1–3), uneven terrain disrupts periodicity and costs considerably more energy (4, 5). Steps onto ground of uneven height perturb forward walking and cause walking speed to fluctuate (Figure 1B), in a pattern dictated by a combination of the terrain and the human compensatory control strategy (Figure 1C). Any number of control strategies are possible, and could be determined by criteria such as to maintain constant speed or minimize energy expenditure. But such control is also potentially complex, because the optimal trajectory could depend on the entire terrain profile, which must therefore be anticipated and planned for at once. It is unknown whether humans anticipate and plan for uneven steps, whether for energy economy or other criteria, and how they manage the complexity. Walking on uneven terrain may thus be indicative of human capacity to anticipate and plan complex control tasks.

Many environmental challenges to walking are anticipated ahead of time. Humans obviously look far ahead to navigate and to plan an overall route. They also look more closely—only a few steps

ahead—to step over an obstacle (6) or to select a foothold (7, 8). But in addition to kinematics, walking also has significant inertial dynamics (9) and thus momentum. For example, birds appear to use dynamics and leg posture to adjust for a change in ground height while walking (10). Similarly, humans appear to compensate for an upward step such as a sidewalk curb by gradually speeding up a few steps ahead, losing momentum while ascending the curb, and finally gradually regaining speed a few steps after (11). A nearly opposite one applies to stepping down a curb, and the compensations of birds and humans seem compatible with energy minimization. It appears that a single uneven step is anticipated ahead of time, and that the compensation takes place over several surrounding steps.

A sequence of multiple uneven steps may require more complex compensation (Figure 1A). The momentum of one step carries forth into succeeding ones, and thus the compensation for one isolated uneven step might be incompatible with its neighbors. It may therefore be important to anticipate and plan a trajectory for many steps at once (Figure 1B), but for computational complexity that typically increases exponentially with the number of steps, termed a "curse of dimensionality" (12, 13). There is thus need for either a means to plan optimal trajectories despite high dimensionality, or a heuristic and practical but suboptimal compensation strategy.

These possibilities depend critically on the mechanics and energetics of walking. One of the key models of walking treats the stance leg like an inverted pendulum, which carries the body center of mass (COM, near the pelvis) in a pendular arc (14). That motion is relatively conservative of mechanical energy, except during transition from one pendulum-like leg to the next, or *step-to-step transition* (15). There is a dissipative collision between leading leg and ground, which is restored by active work by muscles of the trailing leg, at a proportional cost that accounts for most of the metabolic energy expended for level walking (3). Such a model can predict the optimal forward speed fluctuations for ascending the sidewalk curb mentioned above (16). The dissipation also causes one step's forward momentum to decay within only a few consecutive steps, termed a persistence distance (16). It may thus be unnecessary to plan an unlimited number of neighboring steps at once, but rather only a finite, receding horizon of upcoming steps. Indeed, robotic locomotion has long used a similar paradigm of model predictive control (17), whereby optimization is performed for a finite horizon (18), and re-computed each step within a fast timing loop to act as a form of feedback control. Planning over a finite receding horizon is often practically quite similar to truly optimizing over all steps.

The optimization approach has been applied to a variety of human motions. Upper extremity reaching motions have long been thought to minimize kinematic objectives (19), and more recently energy expenditure (20, 21). Similarly, walking routes have been proposed to minimize kinematic objectives such as speed-curvature trade-offs (22), and empirical energy objectives to predict curved walking paths through doorways (23). These results suggest the potential to mechanistically predict the dynamics of walking, but most studies to date have been concerned with steady, periodic gait (24). The challenge of walking on uneven terrain suggests the need to integrate mechanics, energetics, and planning, while somehow avoiding or mitigating high complexity.

The purpose of the present study was to test whether humans can and do optimize their compensations for uneven terrain consisting of a complex pattern of unequal height steps. We used optimization of the pendulum-like walking model to predict speed fluctuation trajectories for a variety of uneven terrain profiles (Figure 1B). We also considered whether a finite planning horizon can yield similar predictions, but without the complexity of a full horizon. We then performed a human subjects experiment, to test whether the optimal compensations could predict human responses to a similar variety of terrains. If humans do favor energy economy as modeled by step-to-step transition dynamics, and the optimization is feasible and predictive for multiple steps, then the model should reasonably predict human responses. This will reveal whether humans favor energy economy when walking on uneven terrain, and whether and how far they plan into the future.

## Results

### Model Predictions

The simple walking model demonstrates compensatory control strategies for walking on uneven terrain. The model shows how optimization criteria predict specific trajectories of walking speed, time duration, and energy cost. Here the proposed *Minimum energy objective* is contrasted with two examples of alternative criteria. These criteria, applied to multiple uneven terrain profiles, result in distinct, terrain-specific predictions, which are to be tested experimentally. Finally, the model predicts how planning horizon affects walking motions, to contrast full vs. finite horizon planning.

Three alternative control strategies are examined: Reactive control, Tight regulation, and the proposed Minimum energy objective. The task is to traverse multiple steps of uneven height in nominal time (Fig. 2A; nominal defined as steady level walking), using a model of pendulum-like leg dynamics (Fig. 2B), where momentum is controlled by modulating active push-off during on each uneven step (Fig. 2C, push-off "PO"). The control strategies are as follows. Reactive control does not anticipate the upcoming terrain, and instead compensates for disturbances after they happen, with the minimum work achievable for this case. Tight regulation is a high gain compensation that looks only one step ahead, and immediately restores speed for that step, regardless of energetic cost. Minimum energy plans an optimal strategy for the entire (full horizon) terrain sequence in advance, to minimize the push-off work for step-to-step transitions. As demonstration, these strategies are applied to a sample uneven terrain profile that gently slopes up and then down over several steps (termed Pyramid, Fig. 2D).

The model predicts that all three strategies can traverse the terrain in nominal overall speed and time (zero cumulative time gain) after traversing the Pyramid terrain, but at different costs. Reactive control and Tight regulation respectively perform 12.1% and 9.2% more push-off work than nominal level gait (0.718 $MgL$ overall work; $MgL$ is the gravitational potential energy of the model's center of mass, COM). The Minimum energy strategy is most economical, performing about 6.2% more work (0.044 $MgL$) than nominal. The optimal speed profile starts several steps before, and ends several steps after the uneven terrain. It achieves economy by pushing off harder and speeding up several steps ahead, in anticipation of a loss of speed while ascending the upward steps. The optimal descent is nearly the reverse in time, regaining speed on the downward steps, and then slowing down again after them. The overall strategy gains time in the first half and then loses it in the second (Fig. 2, Time gain). This saves work compared to any other alternative, and requires 80% less work than the gravitational potential energy of the Pyramid itself (0.225 $MgL$). It is thus possible to ascend steps with less work than would be needed to lift the body against gravity alone. This demonstrates the advantage of planning strategically for an entire terrain sequence at once.

Using the Minimum-energy strategy, the model also predicts distinct compensation strategies for each of seven experimental terrain profiles (Fig. 3), in comparison to level nominal walking (Control). The terrains consist of sequences of discrete ground-height changes, each up to 0.075 $L$ (leg length $L$). The profiles have between one and sixteen such height changes, and are termed Up-step (U), Down-step (D), Pyramid (P, see also Fig. 2), Up-Down (UD), Down and Up-Down (D&UD), Complex 1 (C1), and Complex 2 (C2). The task is to traverse the terrain and regain nominal speed and time, with compensations starting no earlier than six steps before the first uneven step and ending no later than six steps after. Given a full planning horizon, the optimal strategy for Up-step (Down-step) is a tri-phasic speed fluctuation pattern, as described previously (16). The Pyramid strategy (repeated from Fig. 2) is accompanied by parameter variations (Fig. 3, Parameters inset) for three different speeds and four different step lengths (shorter, nominal, and longer fixed lengths, and lengths increasing with speed according to the human preferred relationship, as detailed in Methods). The speed profiles vary in overall amplitude and time for these cases, but all retain a similar scalable shape in terms of discrete steps. The remaining terrains also yield deterministic, scalable, and distinct optimal profiles, whose fluctuation patterns are to be tested against human.

The Minimum-energy strategies (Fig. 3) constitute the main testable predictions for humans, who regardless of self-selected speed and step length, are expected to produce speed fluctuation patterns similar to model. The predictions extend beyond the section of uneven terrain alone, and also include anticipatory speed changes before the first uneven step, and recovery beyond the final uneven step. There are, of course, many other models and optimization objectives possible, but each would be expected to produce distinct predictions for various terrain profiles, making the present model falsifiable.

The model also predicts compensation strategies for shorter, finite planning horizons (Fig. 4). In contrast to the previous full horizon, here the optimization plans based only on horizons of $m$ upcoming steps at time. Only the first step of the $m$-step plan is executed, because the horizon is continually updated and a new compensation strategy planned, based on this finite receding horizon. The planner is aware of the ultimate goal of regaining nominal time, but not of the distant terrain until it enters the horizon. Very short horizons yield speed fluctuations and push-off work trajectories (Fig. 4A and 4B, respectively) very different from full horizon control. Short horizons are less optimal and expend more work, whereas longer horizons are less costly and asymptotically approach the full-horizon optimum (Fig. 4C). There is at least 90% correlation for $m$ of eight or more steps, and thus diminishing advantage to be gained from planning over longer horizons.

## Experimental results

Human subjects produced a unique speed fluctuation profile as they walked on each terrain (Fig. 5). Even though each person walked at their own self-selected overall speed, they exhibited similar patterns in speed fluctuations (Fig. 5, compare individual speeds and overall speed fluctuations). These patterns compared reasonably well with model predictions. We first report a basic summary of the control condition of level walking, which serves as a basis of comparison for the uneven terrains (Fig. 6). There was a small range of self-selected control speeds (see Figure 6A), 1.38 ± 0.10 m/s, 1.52 ± 0.13 m/s, and 1.51 ± 0.12 m/s (mean ± s.d. across subjects) in the control conditions for each of three sets of terrain (see Methods). Each person walked fairly consistently, with average speed varying about 2.4-4.8% c.v. (coefficient of variation) across their control trials. This consistency occurred despite the fact that subjects receiving no feedback regarding walking durations or speeds. Speed also varied somewhat within each control trial, with about 0.034 ± 0.006 m/s root-mean-square variability, or 2.2% c.v.

*1. Humans approximately conserved overall walking speed and duration on uneven terrain*

Subjects walked at similar overall speeds on all uneven terrains regardless of complexity (Fig. 6A). The overall speed and walking duration differences within each set of terrain conditions including corresponding level Control were typically less than 6% and none were statistically significant: Repeated measures ANOVA yielded (experiment set 1) $P = 0.63$ for speed, $P = 0.97$ for segment duration; (set 2) $P = 0.51$ for speed, $P = 0.78$ for segment duration; (set 3) $P = 0.96$ for speed, $P = 0.52$ for segment duration. Each individual's walking speeds were also fairly consistent across trials of a specific terrain, varying by $2 - 3\%$ (coefficient of variation, c.v.). A time loss or difference would be expected if there were no compensation (Fig. 2). The observed conservation of the overall walking duration and walking speed, regardless of the terrain, indicates that humans compensated for all terrains. The approximate time conservation was in accordance with experimental instructions, and despite no feedback having been provided regarding speed or duration at any point in the experiment.

*2. Humans produced a repeatable speed fluctuation pattern for each uneven terrain*

Each terrain yielded a specific speed fluctuation profile vs. time (Fig. 5). Two basic quantifications are the overall speed variability for a terrain (Fig. 6B, root-mean-square variability) and the maximum speed fluctuations within that terrain (Fig. 6C). The variability was greater than corresponding control (all P < 0.05; paired t-tests), ranging about 2 to 5% (c.v.). The maximum fluctuations differed in magnitude and location for different terrains. For example, the largest deviation was +5.68% at $i = -1$ for Up-step terrain, and +4.07% at $i = 1$ and +4.4 at $i = 2$ for Down-step. For the other terrains, the greatest deviations occurred at other locations: $i = 1$ for Up-Down (-4.97%), $i = 3$ for Down-and-Up-Down (-11.08%), $i = 9$ for Pyramid (+7.79%), $i = 16$ for Complex 1 (+7.31%), $i = 16$ for Complex 2 (+7.14%). These maximum speed fluctuations were all greater (all P < 0.05; paired t-tests) than those in the Control conditions (maximum magnitudes 0.86%, 0.89%, 1.03% for the three sets of terrain, respectively). On average, the maximum speed fluctuations were about 7.6 times greater in magnitude on uneven terrain than Control. The uneven terrain speed fluctuations do not support the Tight regulation hypotheses, for which very little deviations were expected.

The compensations included anticipatory and recovery components. Anticipation may broadly be summarized as a tendency to speed up before a first upward step (U, UD, P, C1, C2 terrains), and to slow down before a first downward step (D and D&UD terrains). For initial upward steps, the preceding step ($i = -1$) appeared to speed up 3.74 – 5.70% relative to nominal speed, and the step preceding initial downward steps to slow down about 2.05%. Moreover, the two preceding steps (average $v_i$ for $i = -2, -1$) for uneven terrains tended to exhibit significant anticipatory speed-up or slow-down consistent with prediction (Fig. 6D, all terrains except D&UD P < 0.007; paired t-tests with Bonferroni correction). Similarly, there was also a significant recovery component, summarized by the average speeds of the two steps after uneven terrain (Fig. 6E, all P < 0.007; paired t-tests with Bonferroni correction). The anticipatory adjustments do not support the Reactive compensation and Tight regulation hypotheses, and the recovery adjustments do not support Tight regulation.

The individual speed responses were similar across subjects (Fig. 6F). This is quantified by the correlation between individual speed fluctuations and the (across-subject) average patterns, which were significant and positive for all terrains. The correlations $\rho$ ranged 0.66 ± 0.13 to 0.85 ± 0.11 (for Complex 2 terrain and Up terrain, respectively; mean ± s.d.). Some control trials also had a small amount of correlation (0.46 ± 0.33, 0.32 ± 0.32, 0.29 ± 0.40 respectively for the three sets), although the maximum speed fluctuations were generally about one-sixth the magnitude for uneven terrain (Fig. 6C). Incidentally, compensation strategies that were predicted to be nearly opposite each other were found to be negatively correlated in data. For example, the individual responses for Up- and Down-steps were negatively correlated with average Down- and Up-steps, respectively ( -0.36 ± 0.17 and -0.27 ± 0.29), and similarly for Up-Down and Down-and-Up-Down (0.52 ± 0.083 and -0.53 ± 0.15, respectively). The similarity between individuals across all terrains is indicative of systematically repeatable responses.

*3. Humans walking speed fluctuations were consistent with minimum-energy predictions*

The compensation strategies agreed reasonably well with model predictions for minimizing energy expenditure over a full horizon (Fig 7). Visual inspection reveals an overall resemblance in speed profiles between human and Minimum energy model, particularly when plotted against each other (Fig. 8A). The agreement is quantified by the correlation between experimental fluctuations for a terrain and the corresponding model prediction (13 – 28 steps per terrain for each subject across seven terrains, at least 11 subjects per terrain). Zero correlation would be expected if the model were not predictive or human strategies were random. Instead, the experimental correlation coefficients ranged from 0.35 to 0.67 across the seven terrains (Figs. 7, 8A & B), all of them statistically significant (P-values by terrain: U 1.5e-12, D 1.7e-21, UD 8.9e-19, D&UD 1.3e-22, P 5.7e-21, C1 1.5e-10, C2 6.2e-18). The lowest correlations were for the two Complex terrains (0.35 ± 0.10 for C1 and 0.46 ± 0.09 for C2, mean ± 95% CI), whereas higher correlations were observed for the shorter and simpler terrains. The average correlation across terrains was 0.55 ± 0.11 s.d. Log likelihood ratios for optimal model compared to random models were (mean ± s.d. across randomly shuffled models) U 33 ± 14, D 55 ± 14, UD 50 ± 15, D&UD 55 ± 15, P 81 ± 16, C1 51 ± 15, C2 72 ± 16. This was equivalent to a range of 2.6 – 6.1 bits per step of predictive information.

(For reference, a model predicting only the direction of speed change each step would have 1 bit/step of information.) The minimum Bayes factor was 4.9e14, for U terrain; it is the factor by which a prior odds ratio (proposed model to random model) is adjusted to yield a posterior odds ratio. The correlations and highly significant P-values support the hypothesis that the model is predictive of experimental speed fluctuations, and the log likelihood ratios show that the model adds substantial predictive information.

There were some steps and some terrains where human and model did not agree well. By visual inspection of Up- and Down-steps (Fig. 7), humans usually spent an extra step (about $i = 2$) at greatly deviated speed compared to the model. On Down-and-Up-Down, humans were slower than model on step $i = 4$. And in the middle of both Complex terrains, human speed fluctuations did not vary as much as model (e.g. $i = 8$); they were similar to applying a low-pass filter to model responses. Such differences emphasize that the general agreement between human and model did not apply to every step of every terrain.

*4. A finite planning horizon predicts most human compensations*

The human speed fluctuations could also be predicted reasonably well with a finite planning horizon. This was quantified by the correlation between human responses and model predictions for a range of $m$-step horizons (Fig. 8C). For all terrains, short planning horizons had almost no predictive ability, and longer horizons generally did better. The overall correlation, averaged across terrains (Fig. 8C, gray line) resembled a saturating exponential, with a nearly monotonic increase with horizon length $m$, maximized with the full horizon. The saturation is consistent with model dynamics (Fig. 4), for which the immediate upcoming step is most important, and succeeding steps exponentially less so. This makes it highly advantageous to look at least a few steps ahead, but with diminishing returns for farther lookahead. The observed advantage of longer planning horizons was terrain-specific, in that several (U, UD, D&UD, C1) had peak correlation with a model planning for as few as 5 to 8 steps, and others (D, P, C2) only for a full horizon. As a basic summary, the average correlation was within 92-93% of the saturation value with horizons of six to eight steps. A finite planning horizon of at least such length could thus be regarded as sufficient to predict most human responses.

## Discussion

We had sought to determine whether humans compensate for uneven terrain by planning and adjusting their forward momentum. The human data showed systematic speed fluctuations that were consistent across individuals, and specific to each of the terrain patterns, including complex patterns of consecutive uneven steps. The speed fluctuations included an anticipatory component prior to actually contacting the uneven terrain and showed patterns that were consistent with the minimum-energy optimization model. That optimization becomes increasingly complex for more uneven steps. However, we found it sufficient to consider only a finite horizon of upcoming steps to yield near-optimal economy. We interpret these results to suggest that humans optimally plan and control for uneven terrain, in a manner that is bounded in planning complexity.

The human speed fluctuation patterns were not mere noise. The fluctuations might superficially seem to have little organization (e.g., Fig 5 complex terrain C1), but quantification shows that the patterns were repeatable across subjects, unique to each terrain, and predictable by model. Part of the repeatability may be explained by inverted pendulum dynamics, for example downward steps should gain (and upward steps should lose) forward speed and momentum. But dynamics alone do not explain how walking durations could be conserved across different terrains. Without any compensatory control for terrain variations (i.e. using nominal push-off for all steps), overall walking speeds would have been reduced (16). With compensatory but purely reactive control, it would be possible to regain lost time (Fig. 2, Reactive compensation), but not before losing it. Another possibility would have been to tightly regulate speed to nearly constant with each uneven step (Fig. 2, Tight regulation). But the significant and repeatable speed fluctuations observed (Fig. 6C) do not support these possibilities. Instead, there was clear anticipatory compensation, consistent with the model. There were systematic adjustments before the first uneven step, where subjects sped up significantly before a first upward step and slowed down before a first downward step (Fig. 6D, compare D and D&UD against other terrains). There was also a systematic recovery component beyond the final uneven step (Fig. 6E, compare D and D&UD against other terrains), as predicted by model. Overall speed was conserved by conclusion of the recovery, which is suggestive that the entire trajectory was planned. Our interpretation is that humans plan over multiple steps, perhaps extending from several steps before to several steps after a segment of uneven terrain. Our subjects approximately conserved speed and duration (Fig. 6A), and in part by adjusting their speed ahead of time (Fig. 6D).

Such planning appears to be performed over a multi-step horizon. A simple indication of look-ahead is that walking speed started deviating from steady state ahead of the first uneven step (Fig. 5, Fig. 6D). A better indicator is the predictability of human responses with the model's horizon length $m$ (Fig. 8C). Longer planning horizons generally allowed the model to better predict human (Fig. 8C, gray line) with a gradually saturating behavior. The advantage of planning ahead is explained by simple governing principles, namely the step-to-step transitions of pendulum-like walking, which dissipate energy with each ground collision. About 70% of the forward momentum from one step is carried into the next, and less than 1% of the forward kinetic energy persists beyond seven steps (25), also described as a persistence distance (16). A momentary perturbation, whether from the ground or by the person, thus has consequences for succeeding steps, making it generally optimal to plan a sequence of steps at once. The persistence is also limited, so that there is no advantage to planning infinitely into the future. Much of the predictive ability of the finite horizon model (Fig. 8C) comes from the first few steps of look-ahead, and six to eight steps of look-ahead appear sufficient to explain over 90% of the observed human responses.

This does not, however, mean that humans literally optimize for individual step variations. It might be practical to reduce complexity by reasoning about chunks of steps, for example treating Pyramid terrain (Fig. 4) as simply one chunk with a gentle upward followed by downward undulation rather than twelve individual steps (Fig. 8C, P terrain). It may also be more important for the human to attend to overall, low-frequency ups and downs rather than the detailed variations. This may explain why human responses on complex terrains seemed to be low-pass filtered versions of model (see Results). It is quite possible that the long look-ahead observed here is better represented as a smaller number of chunks or low-frequency step combinations.

The proposed criterion for this planning is energy economy. Much of the natural world imposes unsteady and uneven conditions, making it important to economize not only for the steady, level case, but also for when energy costs are highest. We used a mechanistic and quantitative model to predict how economy may be achieved, as governed step-to-step transitions (16). These dynamics describe how speed is lost with an upward step and gained by a downward one, and how push-off may be modulated to change speed and affect the collision loss. Step-to-step transitions have previously been tested against humans during steady walking, where parameters such as step length or width (15) are readily manipulated and the associated energy cost tested. Such experiments do not apply to unsteady conditions, and so here we experimentally manipulated only the terrain and computationally predicted the compensatory trajectories. And through testing on seven different terrains, there is low probability that a random model could predict human responses well by chance.

Our results suggest that humans reason about energetics, dynamics, and timing for locomotion. Energetics refers to the ability to judge the upcoming terrain profile with respect to the hypothesized criterion of energy minimization. Although humans are understood to prioritize economy for steady, level walking (26), some form of energetic prediction is at least as important for selecting from the many options available for uneven terrain. Dynamics refers to the translation of that criterion into an appropriate sequence of control actions. Humans seem to reason about the momentum lost or gained by a change in ground height, and how active push-off and other control can influence that change. Just as the ability to catch a ball suggests reasoning about the ball's dynamics in flight, the ability to conserve time and energy on uneven terrain implies reasoning about the body's own dynamics, described here by a model of pendulum dynamics and the step-to-step transition. Timing refers to an ability to form an expectation of the time to traverse a given distance (Fig. 1), and to use that to guide the dynamic control actions with minimum energy expenditure. We treat the overall reasoning as tantamount to a central nervous system internal model (27) of walking that enables planning for economy.

The present study is concerned with a type of control that is intermediary between lower- and higher-level concerns. At the lower level, the central nervous system performs control in real time with relatively fast feedback (within tens of milliseconds) from somatosensory and other inputs, much of it mediated at the level of the spinal cord (28). At a higher level, humans can consciously plan many seconds or minutes into the future, for example the best route to the supermarket, the amount of food to be carried on a trek, or whether to walk at all. Much of that planning requires cognition and need not consider step-to-step momentum. The anticipatory, dynamic planning considered here is intermediate, with an apparent temporal update rate on the order a walking step, or about half a second. Spatially, it integrates higher-level terrain awareness with lower-level walking control, and appears to work subconsciously, in that subjects exhibit little cognitive awareness of what they are planning or how their momentum varies on uneven terrain. The planning observed here is reminiscent of upper extremity reaching movements, which are thought to be mediated by internal dynamical models represented in the cerebellum and motor cortex (29, 30). We speculate that similar neural internal models, with short-term storage or working space for anticipatory adjustments, are employed for dynamic planning of locomotion.

The proposed planning horizon may seem longer than where people usually look. Humans allocate most of their gaze two to four steps ahead on rocky, rough terrain (7, 31). Such information is especially important for foot placement, which impacts balance (32–34) and energy expenditure (35). The nearest steps are considered critical for not only humans (6, 36) and robots (37), but also our simple models (16, 32). But on less challenging terrain, humans walk faster, look ahead farther (31), and expend less energy (4, 38). The present terrain is in that category, consisting of broad and flat surfaces that placed little demand for foot placement and perhaps visual attention. Our interpretation is that humans may also look ahead a variety of distances, including far ahead to determine and anticipate a path (36, 39), and intermediate distances (say, at least six to eight steps ahead) to plan forward momentum. Such lookahead might require only a brief glance yet still serve valuable purpose.

The present optimization approach is related to studies of humanoid robot control and simulation. Real-time control for many legged robots is performed with model predictive control (MPC) over a short horizon, computed repeatedly within a timing loop on the order of a millisecond. This acts as a feedback control that achieves stability and near-optimal performance with manageable computations (18, 40, 41), for robots with many more degrees of freedom and more complex contact conditions than modeled here. We employed a slower, per-step timing loop and a longer multi-step planning horizon, because our focus was on economically managing momentum. However, MPC control is generally applicable over a variety of time scales and horizons. Although we designed the control explicitly, it could also be adapted via reinforcement learning techniques, which can learn quite robust control over uneven terrain, typically also with a finite horizon terrain profile obtained through vision and mapped directly to control commands (42, 43). Thus, humans could potentially embed the optimization within a vision-to-control mapping, rather than continually optimize for each step. An interesting observation is that robustness may be achieved with simple rewards such as to make forward progress, without need for explicit stability criteria, because successful progress implies stability. Our present model for human momentum planning assumes the existence of an executive goal set at higher levels, and a lower-level control for fast feedback. This planner could potentially achieve both economy and stability by looking several steps ahead to direct the upcoming step-to-step transition. At present, finite horizon optimization or learning is a highly viable approach, for both understanding humans and controlling robots.

This is the first study we are aware of that uses mechanistic principles to predict transient walking on uneven terrain. There is ample evidence that humans direct their gaze (7) and start taking actions (44) only a few steps ahead, but with few operational principles for determining the actions. Our model is mechanistic in that it produces a walking gait governed entirely by physical first principles (e.g., inverted pendulum and the step-to-step transition). There were no free parameters or opportunities to fit model to data. We are unaware of any other models that are similarly principled and can predict compensation for uneven terrain. Perhaps the closest analogy is an empirical energetic cost model (23) that predicts dynamic paths for economically taking turning paths such as through doorways (on level ground). We consider our mechanistic model to be largely compatible with such empirical costs, and suspect that a single, first-principles model might therefore explain both uneven and turning walking trajectories.

There are a number of limitations to this work. In terms of experimentation, we found the model to be significantly but only modestly predictive of humans (e.g., correlation 0.66 on Down terrain, Fig. 7). This was due in part to the noisiness of human walking, as individuals did not behave identically to each other (correlation 0.85 on Up terrain Fig. 6F), let alone to any model. It may be better to apply larger terrain disturbances to exceed the noise floor, but we intentionally examined relatively small disturbances where the inverted pendulum model of walking is most applicable. Another limitation is that our terrain disturbances consisted of discrete and flat steps, whereas actual uneven ground is more continuous and requires additional decisions regarding foot placement and orientation. We also experimentally tested the optimization hypothesis through speed trajectories, but did not test actual mechanical work or metabolic energetics. The task was too long to measure ground reaction forces to estimate work over many steps, and too short to allow for the steady-state measurement of oxygen consumption needed to estimate energetics. These remain avenues for further testing of model and its alternatives.

Other limitations were from the simplicity of the dynamic walking model. The model applies inverted pendulum dynamics as a fundamental feature of walking. Here we observed some cases where the model's rapid speed fluctuations were not matched by human (Fig. 7; see central step of Pyramid, several steps of Complex 1 and 2). We interpret this low-pass filter effect as the human stance leg behaving less like an inverted pendulum, allowing the COM to deviate somewhat from a pendular arc. Additional model features such as a previously-hypothesized energetic cost for rapid force production (45) could potentially explain some of the mismatches observed here. But even if humans approximate an inverted pendulum on flat and slightly uneven ground, that is certainly not the case for larger terrain disturbances or for stair steps, where it is necessary to use the knees. When ascending large steps, the leading stance knee flexes and extends substantially and can perform substantial positive work for climbing. And when descending large steps, the trailing leg does not behave like a (non-inverted) pendulum, perhaps to actively dissipate gravitational potential energy. Once in swing, it might also be actively flexed to allow the swing foot to clear the step. Dynamic walking models have been devised to include knee joints with actuation (46, 47), legs that telescope (48), actuators with elasticity (49). Such models have potential to capture more aspects of human walking, especially larger terrain disturbances than examined here.

There are, of course, many other degrees of freedom present in human. For example, a feature missing here is actuation of the swing leg, which humans might use to modulate step length, for example to line up with an uneven step or ascend or descend it more economically. Our model takes fixed step lengths at fixed cost, where the control of step length may also exact a cost (24). But transient step adjustments deviate from steady-state step lengths (50, 51), at a cost yet to be modeled. Other relevant degrees of freedom that could be included in a dynamic walking model include the ankles (52) and lateral body motions (32, 53, 54), although with less effect on economy than pendulum-like step-to-step transitions. It is, however, not straightforward to devise appropriate optimization objectives to predict behavior for many degrees of freedom. We adopted a highly simplified model in part to avoid fitting to data, in favor of first principles.

We showed that humans plan and control their gait to economically locomote over uneven terrain. They anticipate upcoming steps and can adjust their speed and momentum to conserve energy and time. These actions resemble a dynamic optimization procedure, which has potential for impractically high dimensionality. Fortunately, walking momentum does not persist for long, and so it appears sufficient to plan dynamics for as little as six to eight steps into the future. Economy has long been established as a governing principle for level, steady walking, and our findings suggest that it similarly applies to unsteady and uneven walking of arbitrary distance.

## Materials and Methods

We performed an experiment to test whether humans optimally compensate for uneven terrain. Predictions were obtained from a simple optimal control model of walking, described in detail previously (11, 16). Here we briefly summarize the model, followed by a description of the present study's experiment.

### Model dynamics

The optimal control model determines a compensation strategy for traversing a sequence of uneven-height steps at minimum energetic cost while conserving travel time (Fig. 2A). The task is to walk down a walkway interrupted by uneven terrain, starting and ending from steady walking on flat ground. The model's only energetic cost is from positive mechanical work needed to power walking. The key feature determining the optimal strategy is that work is required to redirect the body center of mass (COM) velocity between pendulum-like steps. This work is performed by pushing off with the trailing leg each step, and the sequence of such push-offs comprises the decision variables for optimal control. Conservation of traversal time is a constraint that the total time be equal to that for steady walking on flat ground alone. The overall time and the walking dynamics governing the momentum of each step, are expressed as constraints in the optimal control problem.

We used a simple dynamic walking model, in which the legs act like rigid simple pendulums (16, 24). The stance leg acts like an inverted pendulum supporting a point-mass pelvis of mass $M$ (Fig. 2B), and the swing leg act like a simple pendulum of infinitesimal mass. The inverted pendulum passively conserves mechanical energy with an exchange between potential kinetic energy, so that upward (downward) steps come at a loss (gain) of speed and time. Forward speed $v$ is sampled at the mid-stance instance of each step, when the inverted pendulum is vertical. The main energetic event is the step-to-step transition when the COM velocity must be redirected from forward-and-downward at the end of one step, to forward-and-upward at the beginning of the next ($v^-$ and $v^+$ respectively in Fig. 2B). As a simplification, we ignore the swing leg's rotational dynamics and energetics (11), and are presently concerned only with its collision with ground, modeled as a perfectly inelastic impulse dissipating COM energy and speed. The losses may be restored by pushing off impulsively from the trailing leg just before the collision. In the present study, active mechanical work is performed by push-off alone, and passive dissipation only occurs with collision (PO and CO, respectively in Fig. 2C), and steady walking is a matter of equal magnitudes of push-off and collision. It will be shown that on uneven terrain (step height $b_i$ in Fig. 2C), it is actually optimal to purposefully modulate push-off with each step, causing walking speed to fluctuate. For brevity, the equations presented here use dimensionless quantities, with $M$, gravity $g$, and leg length $L$ as base units.

The discrete dynamics between steps are as follows (see 25 for additional detail). Each of $N$ steps is indexed $i$, with the first uneven step located at $i = 0$ (Fig. 2A), so that negative $i$ refers to preparatory steps beforehand. The model takes steps that end with pre-transition velocity $v_i^-$ at an inter-leg angle of $2\alpha$. Work is performed by a pre-emptive push-off $u_i$ (in units of mass-normalized work), followed immediately by the heel-strike collision along the leading leg, resulting in post-collision velocity $v_i^+$. From impulse-momentum (55) principles,

$$v_i^+ = v_i^- \cos 2\alpha + \sqrt{2u_i} \sin 2\alpha . \tag{1}$$

Uneven steps are described by heights $b_i$ above nominal, equivalent to an angular disturbance $\delta_i$ in landing configuration. For a given step length $S$, the disturbance depends on the difference between successive step heights,

$$\delta_i = \sin^{-1} \frac{b_i - b_{i-1}}{S} . \tag{2}$$

Using linearized inverted pendulum dynamics, the dimensionless step time $\tau_i$ of step $i$ is

$$\tau_i = \log \frac{\alpha - \delta_{i+1} + \sqrt{(v_i^+)^2 - 2\alpha(\delta_i + \delta_{i+1}) + \delta_{i+1}^2 - \delta_i^2}}{v_i^+ - \alpha - \delta_i} \tag{3}$$

and the velocity at end of stance $v_{i+1}^-$ is

$$v_{i+1}^- = \frac{1}{2} \left( e^{-\tau_i}(v_i^+ + \alpha + \delta_i) + e^{\tau_i}(v_i^+ - \alpha - \delta_i) \right) . \tag{4}$$

These quantities may be converted to mid-stance instances as follows. The forward speed $v_i$ at mid-stance is

$$v_i = \frac{1}{2} \left( e^{-\tau_i'}(v_i^+ + \alpha + \delta_i) + e^{\tau_i'}(v_i^+ - \alpha - \delta_i) \right) . \tag{5}$$

where mid-stance time $\tau_i'$ is

$$\tau_i' = \log \left( \frac{\sqrt{(v_i^+)^2 - \alpha^2 - 2\alpha\delta_i - \delta_i^2}}{v_i^+ - \alpha - \delta_i} \right) \tag{6}$$

Nominal model parameters were selected to correspond to typical human walking. The nominal gait was for a person with leg length $L$ of 1 m walking at 1.5 m/s, with step length of 0.79 m and step time of 0.53 s. Here we constrained the model to take steps of fixed length, and examined alternative step lengths in parameter sensitivity studies. These included shorter and longer steps (0.59 and 0.96 m, respectively), as well as preferred step lengths increasing with speed according to the preferred human relationship, approximately $v^{0.42}$ (24, 56). The nominal parameter values were $\alpha = 0.41$, push-off $0.0342\ MgL$, step time $1.665\ g^{-0.5}L^{0.5}$, pre-collision speed $0.601\ g^{0.5}L^{0.5}$, and mid-stance speed $0.44\ g^{0.5}L^{0.5}$. Most step heights were in increments of $b = 0.075L$, equivalent to about 7.5 cm for a human. Previous parameter studies with this model (16, 25) have shown that similar fluctuation patterns across a wide range of parameter values.

The energetics of the present model only include push-off work. This is motivated by the observation that the human COM moves like an inverted pendulum each step, accompanied by an exchange of kinetic with gravitational potential energy (57). Although a pendulum conserves energy, the step-to-step transition requires mechanical work. It predicts how humans perform increasing positive mechanical work per step with variables such as walking speed and step length (15, 58), along with an approximately proportional contribution to metabolic cost. These mechanistic dynamics also produce a periodic walking gait, as also demonstrated by walking robots (59). Of course, humans have many more degrees of freedom and many muscles to control them, which could result in energetics very different from model. Even in the simple dynamic walking models, there are other costs such as for moving the legs back and forth (24) or adjusting foot placement (32). Here we focus on the fundamentals of step-to-step transitions, which are hypothesized to explain most of the human energetic cost (3, 15). The model does not have enough parameters to allow for arbitrary fitting to data and is to be tested in its ability to predict distinct speed adjustments for multiple terrains.

# Optimal control problem formulation

The optimal control problem is to produce a sequence of push-offs to negotiate a series of uneven steps with minimum work, and with no loss of overall speed compared to steady level waking. The sequence consists of $N$ push-offs $u_i$ where the control is exerted for each step $i$. The problem starts and ends with nominal walking at speed $V$, and takes the same amount of overall time as $N$ steps of nominal walking with nominal step time $T$, despite a middle interval of uneven steps. Also serving as a constraint are the model dynamics, including pendulum-like walking punctuated by the step-to-step transition. The optimization may be formulated as:

> Minimize total work $\quad \sum_{i=1}^{N} u_i$
> Subject to
> > Nominal speed at beginning and end: $\quad v_{i_0} = V, v_{i_N} = V$
> > and Nominal total time: $\sum_i \tau_i = T \cdot N$
> > and Walking dynamics: $f(v_{i+1}, v_i, \tau_i, u_i) = 0$

The walking dynamics describe how the forward speed of the next step $v_{i+1}$ and the intervening step duration $\tau_i$ (by combining Eqns. 1 − 6) are related to the current step's speed and push-off ($v_i$ and $u_i$). The steps are indexed such that the zero index $i = 0$ is the first uneven step, $i_0$ is the level, nominal step at the beginning, and $i_N$ the nominal step at the end. Any number of uneven steps occur in the middle, padded on both sides by several level (but not necessarily nominal) steps to allow the model to anticipate and recover from the disturbance. We chose a padding of six steps, for example the longest uneven interval was sixteen steps long (Complex 1 and 2), padded on each side to yield $N = 28$. The six-step advance padding is sufficient to gain most of the economic advantage of planning ahead, which keeps increasing but negligibly so for more steps (16). Padding after the last uneven step serves a different purpose, which is to define an objective terminal goal of regaining nominal walking and timing on level terrain.

The optimization was performed over horizons of various lengths. A full horizon refers to optimizing all $N$ steps until arriving back to nominal walking, yielding a full control trajectory. A finite horizon refers to repeatedly optimizing with knowledge of only $m$ upcoming steps, and nothing beyond that horizon. The objective is to regain nominal timing by the $m$th step (or sooner if the $N$th step overall occurs first). To keep track of nominal timing, the optimization is informed of the cumulative time gain or loss thus far relative to nominal. The finite horizon yields $m$ commands (which are suboptimal with respect of full horizon) and executes only the first one. The optimization is then repeated anew each step, starting from the end of the previous step and still intending to meet the original nominal timing, but over a new $m$-step horizon. This receding horizon is generally suboptimal but can approach the full horizon's optimality with sufficient $m$.

The primary prediction of interest was speed fluctuation patterns. This refers to the waveform shape for $v_i$, as opposed to absolute waveform amplitude. The point-mass model is not expected to accurately predict speed amplitudes, which scale slightly with leg length (or relative terrain amplitude) and self-selected walking speed (16). However, the fluctuation patterns are remain quite similar regardless of parameter values such as average speed, body mass, and leg length (11, 16). The optimization hypothesis is therefore to be tested through scale-invariant correlation of speed fluctuations (Fig. 3).

*Alternative hypotheses: Reactive compensation and Tight regulation*

We considered two alternative ways to compensate for uneven terrain while conserving time. One, termed Reactive compensation, refers to an optimal control that does not act until the terrain has been encountered. It then reactively optimizes a control sequence that will regain time and conserve energy. Even though it is optimal, it is not anticipatory, and is thus suboptimal compared to a control that acts ahead of time. The second alternative, Tight regulation, refers to a feedback controller that maintains constant step time for each step, despite disturbances. It adjusts each push-off to ensure constant time, and thus need not plan ahead. This type of control is simple to perform but is suboptimal because it does not take advantage of past or future information. A prediction from Reactive compensation is that there should be no anticipatory speed fluctuations before the uneven terrain. A prediction from Tight regulation is that there should be very small speed fluctuations across uneven terrain.

## Experiment

We measured speed fluctuations as healthy adult subjects walked on each of seven different terrain profiles plus level control, for a total of eight conditions. The profiles (Fig. 3) consisted of an integer number (one to sixteen) of evenly spaced steps, each deviating in height by a small integer multiple (between -3 and 3) of 2.54 cm. The eight profiles were labeled with the following names: Control, Up-step (U), Down-step (D), Up-Down (UD), Down-and-Up-Down (D&UD), Pyramid (P), Complex 1 (C1), and Complex 2 (C2). Each was assembled from layers of polystyrene insulation foam providing a flat surface for each footfall. Each profile occurred about halfway down a level walkway, with subjects initially walking with a steady, level gait, and also ending with a similar steady, level gait after the terrain. There were two groups of subjects walking in three sets of trials. The first group ($N$ = 12; 7 males, 5 females, all under 30 years age) performed trials of Control, U, D, UD, D&UD. The second group ($N$ = 11; 7 males, 4 females, all under 30 years age) performed trials of Control, P in one set, and Control, C1, C2 in another set. A separate Control was collected for each set of terrains (called n1, n2, n3), and consisted of level ground in the same walkway, but to the side of the uneven steps. The entire walkway was about 30-40 m in length, but data were only analyzed for a middle portion encompassing the terrain plus six steps (about 3.5 m) before and after the terrain. The six-step padding was included to allow walking speed to deviate from nominal both before and after the uneven terrain, as has been observed in both model and human (11, 16).

In all conditions, subjects walked at self-selected, comfortable speed. The main instruction was to walk from a start line and past a finish line "in about the same time" throughout the experiment, to avoid large variations in overall speed across conditions. This instruction was intended to provide only broad context, because the hypotheses were not dependent on any particular speed, and so subjects received no feedback about their timing. There was a brief pause between trials, in which subjects turned around and stood briefly before starting the next trial, in opposite direction. For example, the U and D conditions, and C1 and C2, both consisted of the same terrain in opposite directions. Except for this pairing, the conditions were conducted in random order, with at least four to seven trials per condition in each experimental set (all chosen randomly), also interleaved with occasional Control trials inserted at random within each condition. Prior to data collection, subjects walked several times on Control and uneven terrains to gain familiarity with the walkway and terrains. For the first set of terrains (U, D, UD, D&UD, ordered randomly), there was a visual cue on the floor, a paper sticker placed approximately 5 m from the first uneven step. This was intended to help subjects line up their steps, but we observed that subjects paid little attention to the cue, and it was therefore eliminated for the three remaining terrains (P, C1, C2).

Walking speed and timing was measured inertially, using inertial measurement units (IMUs) fixed atop the instep of each foot. Our interest was in the forward speed of the body each stride, defined as the stride length divided by stride duration for each foot, assuming the body travels the same distance as the feet. Stride measures were found by integrating IMU accelerations to yield instantaneous foot displacement. We used a standard algorithm to fuse sensor data and reduce integration drift (60), where zero-velocity update is performed when the foot is briefly stationary in the middle of stance. We estimated body speed by interleaving each alternating foot's speed for each stride, assigning it to the preceding mid-stance instance in time. There was generally some drift between the two feet, which was reduced by linear de-trending their displacements to ensure agreement on overall distance travelled. The resulting body speed served as basis of comparison against corresponding values calculated from the walking model. Data were analyzed for a central segment of the walkway including uneven terrain plus six steps before and after to capture the deviation from steady walking. To compare between trials and conditions, the time $t = 0$ was assigned to the footfall instant for the first uneven step of any terrain, or next to it for Control. A trajectory of discrete walking speeds was thus measured for each trial, and subject's trials within a condition averaged at the discrete step numbers. The corresponding step timings were also averaged to yield discrete sequence of average speeds and average timings per subject and condition.

We compared each subject's average speed trajectory against all subjects and against model, for each terrain condition. To determine whether subjects were consistent with each other, each subject's trajectory was compared against the average trajectory across subjects, using a Pearson correlation coefficient $\rho$, as a measure of similarity, with one correlation per subject and terrain. To test whether human subjects were consistent with the model, their trajectories were also compared against the model's, also using correlation. Here, all experimental trajectories for a terrain (13 – 28 steps per terrain, a total of 133 steps for each subject) were correlated against the model's, yielding one correlation coefficient per terrain. Such correlations are independent of scale, and thus test the model's predicted fluctuation patterns (16) regardless of an individual fluctuation amplitudes, which may depend on a subject's absolute walking speed, body mass, or body size. A truly predictive model should be positively correlated with human responses, whereas a random model (null hypothesis) would be expected to have zero correlation. Statistical significance of the human vs. model correlation coefficients was tested with a threshold of $P < 0.05$. We also performed a few additional tests of hypothesized planning criteria based on speed data. For the No compensation hypothesis, we expected that average speeds would be lower (or walking durations higher) on uneven than level terrain. For the Tight speed regulation hypothesis, we expected that speeds would fluctuate very little, similar to level. And for the Minimum energy hypothesis, we expected that speeds would fluctuate significantly, including before and after the uneven terrain (termed anticipatory and recovery compensations, defined as two steps preceding or following terrain).

To test the effect of horizon length, we computed the correlations above for models of varying horizon length $m$. Optimal predictions were computed for $m$ ranging up to full horizon, and these were correlated against empirical human responses. If humans plan with a finite horizon, they may exhibit peak correlation with a finite model. If humans plan optimally, they may correlate best with a full horizon model. We used this same approach to also test for sensitivity to shorter (3-step) padding before and after terrain. We found shorter padding to yield negligible difference in results, with the same correlation patterns and no more than 0.05 difference in correlation coefficients compared to the nominal 6-step padding.

We also computed the log likelihood ratio (Bayes factor) for the model relative to random predictions. To facilitate comparison between different individuals with the dimensionless model, each subject's speed fluctuation trajectories were scaled to model's amplitude, using linear regression between model and human. The scaled trajectories across individuals represent a distribution of speed fluctuation trajectories that are potentially predicted by model. The log likelihood of each terrain trajectory was calculated from the summed log-likelihoods across steps, assuming a t-distribution centered at the model prediction for each step. Alternative random (null) hypotheses were generated by shuffling the model's trajectories, thus preserving the distribution of predicted speed fluctuations but not the sequence. A log likelihood ratio for each terrain was computed by subtracting (from model's) log likelihoods from each of 1000 shuffled trajectories, to yield a distribution of log-likelihood ratios (tending toward normal distribution), which were divided by the number of steps for each terrain to yield per-step log likelihoods, reported as mean and standard distribution. These were also converted to base 2 and reported as binary bits of added information per step, relative to an average random model.

## Acknowledgments

This work is supported by NSF DGE 0718128, the ONR ETOWL program, NIH AG030815, the Dr. Benno Nigg Research Chair (University of Calgary), and NSERC (Natural Sciences and Engineering Research Council of Canada) Discovery program and Canada Research Chair (Tier 1) program.

## References

1. M. Y. Zarrugh, F. N. Todd, H. J. Ralston, Optimization of energy expenditure during level walking. *Eur. J. Appl. Physiol.* **33** (1974).
2. H. Elftman, Biomechanics of muscle. *J. Bone Joint Surg.* **48-A**, 363–377 (1966).
3. A. D. Kuo, J. M. Donelan, A. Ruina, Energetic consequences of walking like an inverted pendulum: step-to-step transitions. *Exercise and sport sciences reviews* **33**, 88 (2005).
4. R. G. Soule, R. F. Goldman, Terrain coefficients for energy cost prediction. *J Appl Physiol* **32**, 706–708 (1972).
5. A. S. Voloshina, A. D. Kuo, M. A. Daley, D. P. Ferris, Biomechanics and energetics of walking on uneven terrain. *Journal of Experimental Biology* **216**, 3963–3970 (2013).
6. A. E. Patla, S. D. Prentice, S. Rietdyk, F. Allard, C. Martin, What guides the selection of alternate foot placement during locomotion in humans. *Exp Brain Res* **128**, 441–450 (1999).
7. J. S. Matthis, B. R. Fajen, Humans exploit the biomechanics of bipedal gait during visually guided walking over complex terrain. *Proc. Biol. Sci.* **280**, 20130700 (2013).
8. S. Sukumar, A. A. Ahmed, Walking: How visual exploration informs step choice. *Current Biology* **31**, R376–R378 (2021).
9. A. D. Kuo, J. M. Donelan, Dynamic principles of gait and their clinical implications. *Phys Ther* **90**, 157–174 (2010).
10. A. V. Birn-Jeffery, *et al.*, Don't break a leg: running birds from quail to ostrich prioritise leg safety and economy on uneven terrain. *Journal of Experimental Biology* **217**, 3786–3796 (2014).
11. O. Darici, A. D. Kuo, Humans optimally anticipate and compensate for an uneven step during walking. *eLife* **11**, e65402 (2022).
12. R. Bellman, The theory of dynamic programming. *Bull. Amer. Math. Soc.* **60**, 503–515 (1954).
13. A. E. Bryson, Y.-C. Ho, *Applied Optimal Control* (John Wiley & Sons, 1975) (August 28, 2019).
14. R. M. Alexander, Simple models of human motion. *Applied Mechanics Reviews* **48**, 461–469 (1995).
15. J. M. Donelan, R. Kram, A. D. Kuo, Mechanical work for step-to-step transitions is a major determinant of the metabolic cost of human walking. *Journal of Experimental Biology* **205**, 3717–27 (2002).
16. O. Darici, H. Temeltas, A. D. Kuo, Anticipatory control of momentum for bipedal walking on uneven terrain. *Scientific Reports* **10**, 540 (2020).
17. D. Q. Mayne, Model predictive control: Recent developments and future promise. *Automatica* **50**, 2967–2986 (2014).
18. S. Feng, E. Whitman, X. Xinjilefu, C. G. Atkeson, Optimization-based Full Body Control for the DARPA Robotics Challenge. *Journal of Field Robotics* **32**, 293–312 (2015).
19. C. M. Harris, D. M. Wolpert, Signal-dependent noise determines motor planning. *Nature* **394**, 780–784 (1998).
20. H. J. Huang, R. Kram, A. A. Ahmed, Reduction of metabolic cost during motor learning of arm reaching dynamics. *Journal of Neuroscience* **32**, 2182–2190 (2012).

21. J. D. Wong, T. Cluff, A. D. Kuo, The energetic basis for smooth human arm movements. *eLife* **10**, e68013 (2021).
22. G. Arechavaleta, J.-P. Laumond, H. Hicheur, A. Berthoz, An Optimality Principle Governing Human Walking. *IEEE Transactions on Robotics* **24**, 5–14 (2008).
23. G. L. Brown, N. Seethapathi, M. Srinivasan, A unified energy-optimality criterion predicts human navigation paths and speeds. *PNAS* **118** (2021).
24. A. D. Kuo, A simple model of bipedal walking predicts the preferred speed-step length relationship. *Journal of Biomechanical Engineering* **123**, 264–9 (2001).
25. O. Darici, H. Temeltas, A. D. Kuo, Optimal regulation of bipedal walking speed despite an unexpected bump in the road. *PLOS ONE* **13**, e0204205 (2018).
26. H. J. Ralston, Energy-speed relation and optimal speed during level walking. *Int. Z. Angew. Physiol. Einschl. Arbeitsphysiol.* **17**, 277–283 (1958).
27. E. Todorov, Optimality principles in sensorimotor control. *Nature neuroscience* **7**, 907–915 (2004).
28. D. A. McCrea, Spinal circuitry of sensorimotor control of locomotion. *The Journal of Physiology* **533**, 41–50 (2001).
29. M. Kawato, Internal models for motor control and trajectory planning. *Current Opinion in Neurobiology* **9**, 718–727 (1999).
30. S. H. Scott, J. F. Kalaska, Reaching movements with similar hand paths but different arm orientations. I. Activity of individual cells in motor cortex. *J Neurophysiol* **77**, 826–852 (1997).
31. J. S. Matthis, J. L. Yates, M. M. Hayhoe, Gaze and the Control of Foot Placement When Walking in Natural Terrain. *Current Biology* **28**, 1224-1233.e5 (2018).
32. A. D. Kuo, Stabilization of lateral motion in passive dynamic walking. *International Journal of Robotics Research* **18**, 917–930 (1999).
33. C. E. Bauby, A. D. Kuo, Active control of lateral balance in human walking. *J Biomech* **33**, 1433–1440 (2000).
34. S. M. O'Connor, A. D. Kuo, Direction-dependent control of balance during walking and standing. *J. Neurophysiol* **102**, 1411–1419 (2009).
35. S. M. O'Connor, H. Z. Xu, A. D. Kuo, Energetic cost of walking with increased step variability. *Gait & Posture* **36**, 102–107 (2012).
36. T. Higuchi, Visuomotor Control of Human Adaptive Locomotion: Understanding the Anticipatory Nature. *Frontiers in Psychology* **4** (2013).
37. T. Koolen, T. de Boer, J. Rebula, A. Goswami, J. Pratt, Capturability-based analysis and control of legged locomotion, Part 1: Theory and application to three simple gait models. *The International Journal of Robotics Research* **31**, 1094–1113 (2012).
38. D. B. Kowalsky, J. R. Rebula, L. V. Ojeda, P. G. Adamczyk, A. D. Kuo, Human walking in the real world: Interactions between terrain type, gait parameters, and energy expenditure. *PLOS ONE* **16**, e0228682 (2021).
39. D. Bernardin, *et al.*, Gaze anticipation during human locomotion. *Exp Brain Res* **223**, 65–78 (2012).
40. M. Johnson, *et al.*, Team IHMC's Lessons Learned from the DARPA Robotics Challenge Trials. *Journal of Field Robotics* **32**, 192–208 (2015).
41. S. Kuindersma, *et al.*, Optimization-based locomotion planning, estimation, and control design for the atlas humanoid robot. *Autonomous Robots* **40**, 429–455 (2016).
42. N. Heess, *et al.*, Emergence of Locomotion Behaviours in Rich Environments. *arXiv:1707.02286v2* (2017) (August 28, 2019).

43. X. B. Peng, G. Berseth, K. Yin, M. Van De Panne, DeepLoco: Dynamic Locomotion Skills Using Hierarchical Deep Reinforcement Learning. *ACM Trans. Graph.* **36**, 41:1-41:13 (2017).
44. A. E. Patla, How is human gait controlled by vision. *Ecological Psychology* **10**, 287–302 (1998).
45. J. R. Rebula, A. D. Kuo, The Cost of Leg Forces in Bipedal Locomotion: A Simple Optimization Study. *PLOS ONE* **10**, e0117384 (2015).
46. P. G. Adamczyk, S. H. Collins, A. D. Kuo, The advantages of a rolling foot in human walking. *Journal of Experimental Biology* **209**, 3953–3963 (2006).
47. J. C. Dean, A. D. Kuo, Energetic costs of producing muscle work and force in a cyclical human bouncing task. *J Appl Physiol* **110**, 873–880 (2011).
48. M. Srinivasan, A. Ruina, Computer optimization of a minimal biped model discovers walking and running. *Nature* **439**, 72–5 (2006).
49. Z. Gan, Y. Yesilevskiy, P. Zaytsev, C. D. Remy, All common bipedal gaits emerge from a single passive model. *Journal of The Royal Society Interface* **15**, 20180455 (2018).
50. L. V. Ojeda, J. R. Rebula, A. D. Kuo, P. G. Adamczyk, Influence of contextual task constraints on preferred stride parameters and their variabilities during human walking. *Medical Engineering & Physics* **37**, 929–936 (2015).
51. M. Snaterse, R. Ton, A. D. Kuo, J. M. Donelan, Distinct fast and slow processes contribute to the selection of preferred step frequency during human walking. *Journal of Applied Physiology* **In review** (2010).
52. K. E. Zelik, T.-W. P. Huang, P. G. Adamczyk, A. D. Kuo, The role of series ankle elasticity in bipedal walking. *Journal of Theoretical Biology* **346**, 75–85 (2014).
53. J. M. Donelan, R. Kram, A. D. Kuo, Mechanical and metabolic determinants of the preferred step width in human walking. *Proc. Biol. Sci* **268**, 1985–1992 (2001).
54. J. R. Rebula, L. V. Ojeda, P. G. Adamczyk, A. D. Kuo, The stabilizing properties of foot yaw in human walking. *J Biomech* **53**, 1–8 (2017).
55. A. D. Kuo, Energetics of actively powered locomotion using the simplest walking model. *Journal of Biomechanical Engineering* **124**, 113–20 (2002).
56. D. W. Grieve, Gait patterns and the speed of walking. *Biomedical Engineering* **3**, 119–122 (1968).
57. G. A. Cavagna, F. P. Saibene, R. Margaria, External work in walking. *J Appl Physiol* **18**, 1–9 (1963).
58. P. G. Adamczyk, A. D. Kuo, Redirection of center-of-mass velocity during the step-to-step transition of human walking. *Journal of Experimental Biology* **212**, 2668–2678 (2009).
59. S. Collins, A. Ruina, R. Tedrake, M. Wisse, Efficient bipedal robots based on passive-dynamic walkers. *Science* **307**, 1082–1085 (2005).
60. J. R. Rebula, L. V. Ojeda, P. G. Adamczyk, A. D. Kuo, Measurement of foot placement and its variability with inertial sensors. *Gait Posture* **38**, 974–980 (2013).

# Figures and Tables

**A.** Anticipation of uneven terrain

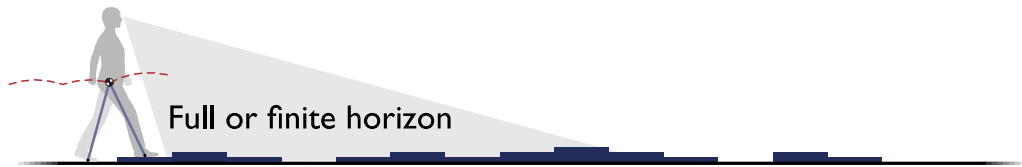

**B.** Uneven terrain trajectories

**C.** Anticipatory planning

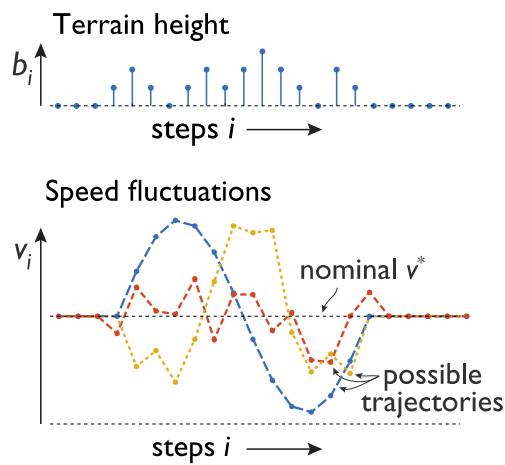
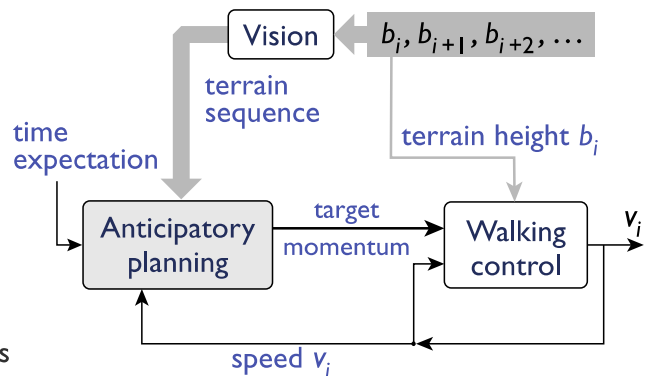

**Figure 1.** Anticipation of uneven terrain over a full or finite horizon. (A.) The human can anticipate upcoming terrain, using vision to observe varying ground height for each step, and plan how momentum should be adjusted for economy or other objectives. The planning could be based on a full horizon of an unlimited number of uneven steps, or a finite horizon of $m$ steps. (B.) Terrain of varying height is a sequence ($b_i$ for steps numbered $i$) of perturbations to walking. For a given terrain, the human produces a trajectory of walking speed ($v_i$) varying with each step. Speed can fluctuate about the nominal level speed ($v^*$), and is influenced by the dynamics of human walking, the terrain, and compensatory control performed by human. A few possible trajectories are shown, dependent on terrain and control strategy. (C.) The human control strategy may include anticipatory planning to determine a dynamic, compensatory plan. The planning criteria could include energy economy or other objectives, subject to a time expectation constraining how much time to take. The plan may be represented as a trajectory of reference speeds ("target momentum") for upcoming steps. We assume that planning is informed by vision of the upcoming terrain sequence, and that the planned trajectory may drive a lower-level walking controller that can produce motor commands based on the target and local feedback of body state and speed. This study tests whether humans produce anticipatory compensations to conserve energy and time, and whether a finite horizon of future steps is sufficient to predict the compensation strategy.

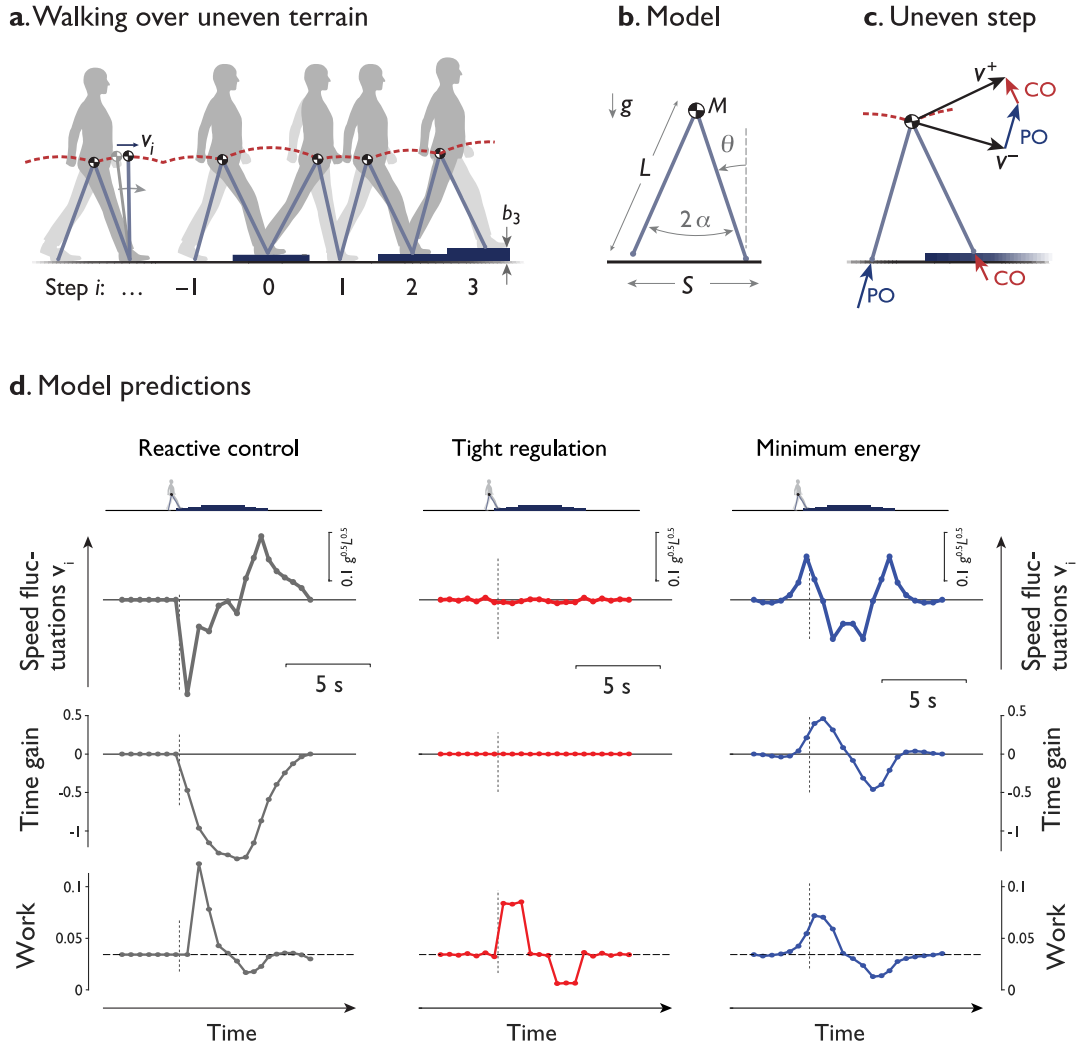

**Figure 2.** Model predictions for hypothetical strategies for walking over uneven terrain. (a.) The task is to traverse successive uneven steps of height $b_i$, while conserving the same overall walking time and speed as nominal level ground. Steps are numbered consecutively with index $i$, where $i = 0$ corresponds to first step deviating from level ground, (b.) Simulations are performed with a simple walking model with pendulum-like legs. Each step, whether even or (c.) uneven, is punctuated by a step-to-step transition, where the trailing leg performs active, impulsive push-off (PO) just prior to the leading leg's dissipative and impulsive collision (CO) with ground. (d.) Model predictions are illustrated for a sample terrain (termed Pyramid) of several steps up and then several down, for three hypothetical strategies: Reactive control that optimally adjusts the walking pattern only after encountering uneven terrain (i.e., no anticipation); Tight time regulation, where each step's timing is kept as constant as possible to reject disturbances; and Minimum energy control, which looks ahead to a full horizon of uneven steps and plans an optimal control including anticipatory adjustments. Strategies are described in terms of (top row:) speed fluctuations vs. time, (middle:) cumulative time gain relative to nominal walking, and (bottom:) push-off work vs. time. All strategies avoid an overall loss of time, but Minimum energy required the least work. Data are sampled discretely (dot symbols) for each step, with speed $v_i$ sampled at mid-stance instant ($i = 0$ denoted by vertical dashed line). In the corresponding experiment, human subjects walked over similar Pyramid terrain in a walkway (30 m long), with step height increments of 7.5 cm. Model parameters: $\alpha = 0.41$, nominal mid-stance speed $V = 0.44$.

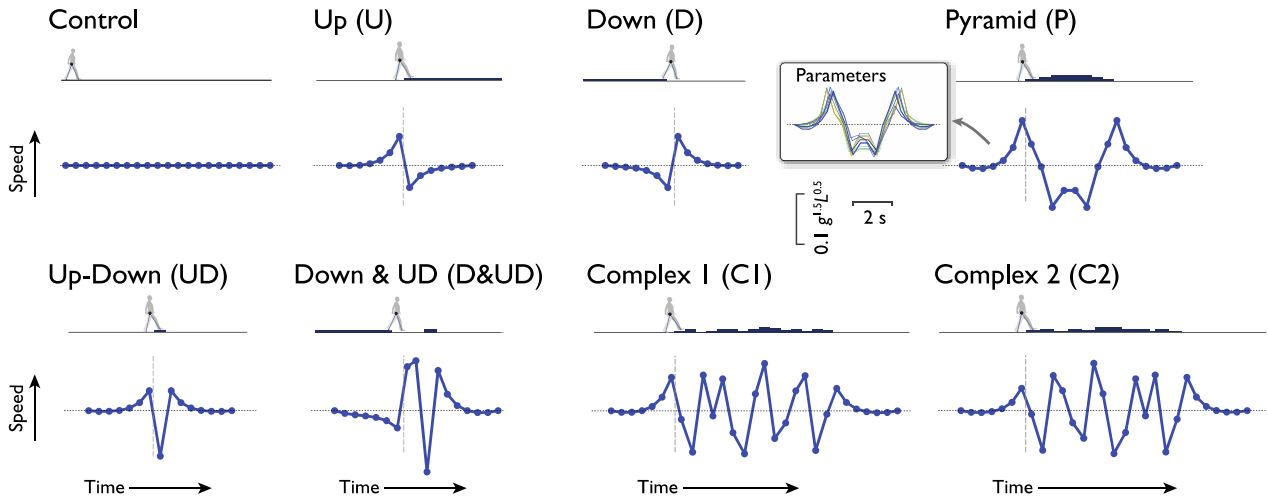

**Figure 3.** Predicted speed fluctuations vs. time for walking on different terrains with minimum energy. Simple model is optimized over a full horizon for eight terrains: Control on level ground, a single Up-step (U), a single Down-step (D), a Pyramid (P) with a flat top, an Up-Down (UD) sequence, a Down-step followed by a level step and an Up-Down sequence (D&UD); Complex terrain 1 (C1); and its reverse, Complex terrain 2. Objective is to perform minimum push-off work while traversing the terrain in same overall duration and speed as nominal level Control. Shown are the minimum-energy speed fluctuations, sampled discretely once per step (filled dots) at model's middle-stance instant. Overall speed (in normalized units of $g^{0.5}L^{0.5}$; gravity $g$, leg length $L$) is equivalent to about 1.5 m/s. Inset shows Parameter variations for Pyramid terrain, with twelve combinations of step lengths (ranging 0.59 m – 0.96 m) and speed (1 – 1.5 m/s), all scaled and superimposed to illustrate similarly shaped speed trajectories. Terrains consisted of discrete changes in step height (up to 7.5 cm each) in sequences of up to sixteen steps.

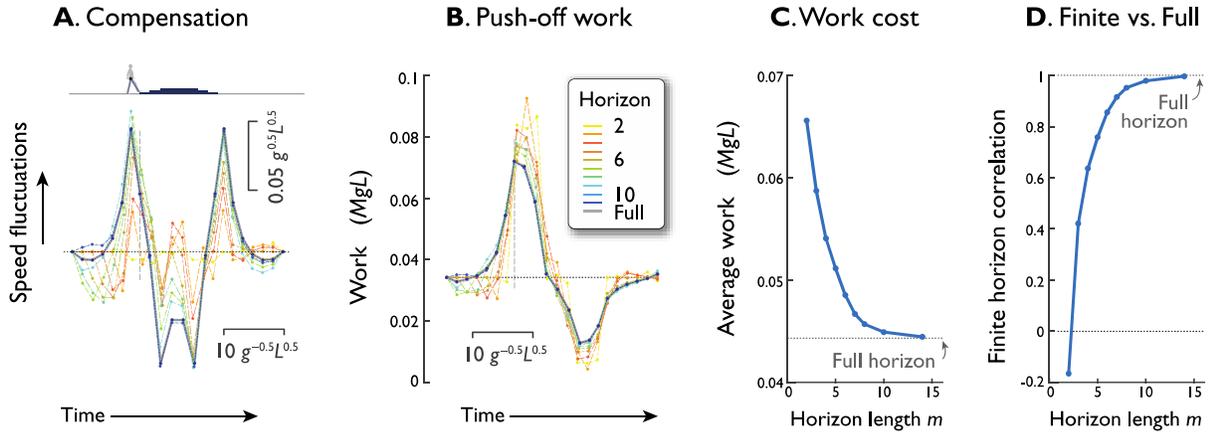

**Figure 4.** Effect of finite planning horizon for model walking over Pyramid (P) terrain with Minimum energy. (A.) Compensation strategy in terms of optimal speed fluctuations vs. time, for finite horizon lengths $m$ ranging 2 to 14 (lighter to darker colored lines). With longer horizons, fluctuation pattern approaches full horizon optimum (gray lines). (B.) Push-off work trajectories vs. time for traversing the terrain in nominal time, for a range of horizon lengths. (C.) Work cost vs. horizon length $m$. Cost is defined as average push-off work per step to traverse the terrain. With longer horizons, cost approaches full horizon optimum (Fig. 3). (D.) Correlation of finite horizon compensation strategies with full horizon optimization. Correlation is between speed fluctuations for each finite horizon, and a full horizon (defined as 21 steps here). For all plots, the longer the horizon, the greater the resemblance to planning over a full horizon. As few as seven to eight steps are sufficient to gain most of the economy and performance of a full horizon. Optimization minimizes push-off work to traverse terrain while conserving overall speed. The conditions are equivalent to human walking at 1.5 m/s with 7.5 cm increments in step height on Pyramid terrain.

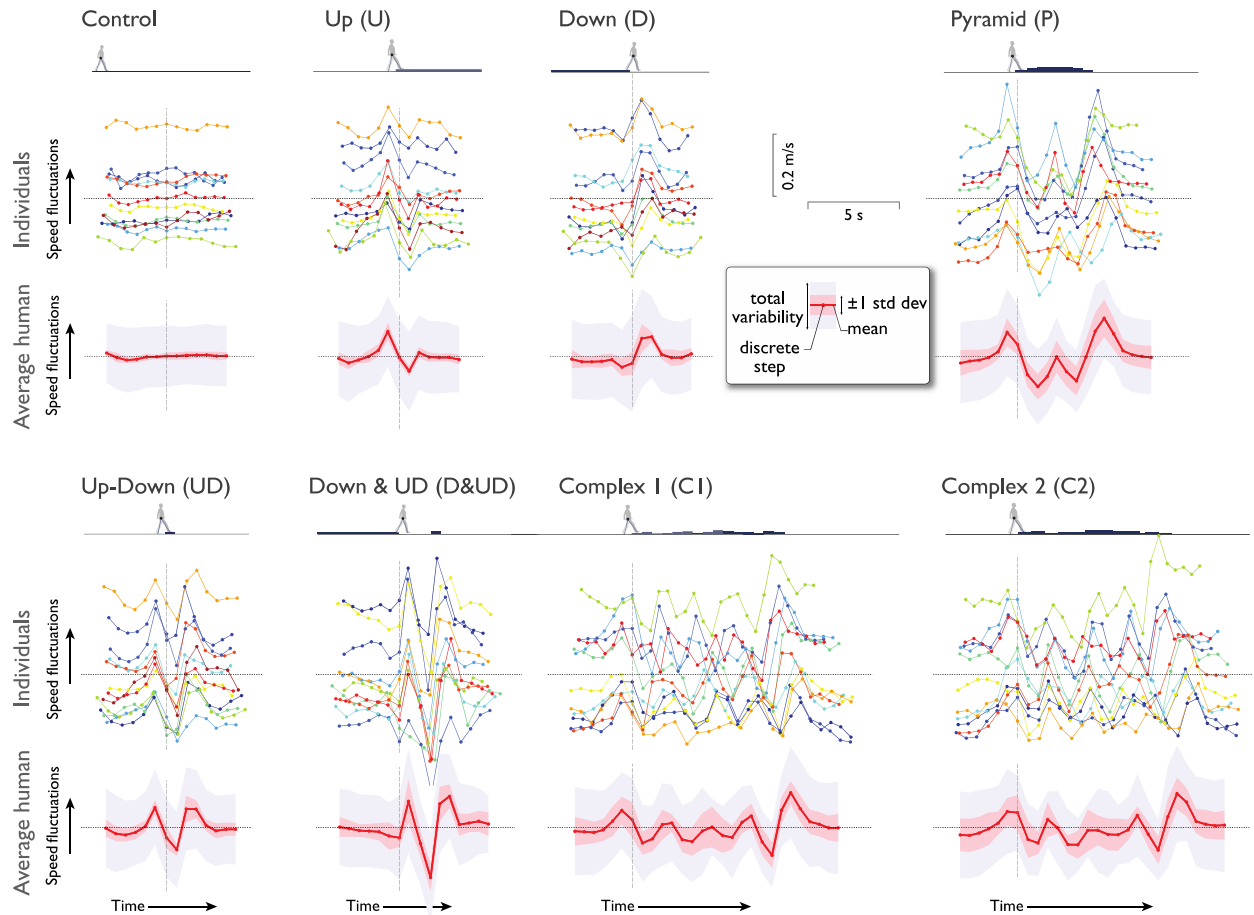

**Figure 5:** Human walking speed fluctuations vs. time on uneven terrain. Terrains are level Control, Up-step (U), Down-step (D), Pyramid (P), Up-Down (UD), Down-and-up-down (D&UD), Complex terrain 1 (C1), Complex terrain 2 (C2). Plots are arranged in pairs of (top:) each individual subject's speed fluctuations and (bottom:) overall speed fluctuations for all subjects. Two types of variability are shown: total variability (light shaded areas) based on ±1 standard deviation across all individuals and all absolute speeds, and ±1 standard deviation of fluctuations (darker shaded areas) from each individual's mean speed. Individual subject traces are means across all of each subject's trials. Speed is defined as step length divided by step time, assigned to the discrete middle-stance instant of each step (indicated by dot symbols). All plots are aligned in time to the middle-stance instant of the first step on the uneven terrain.

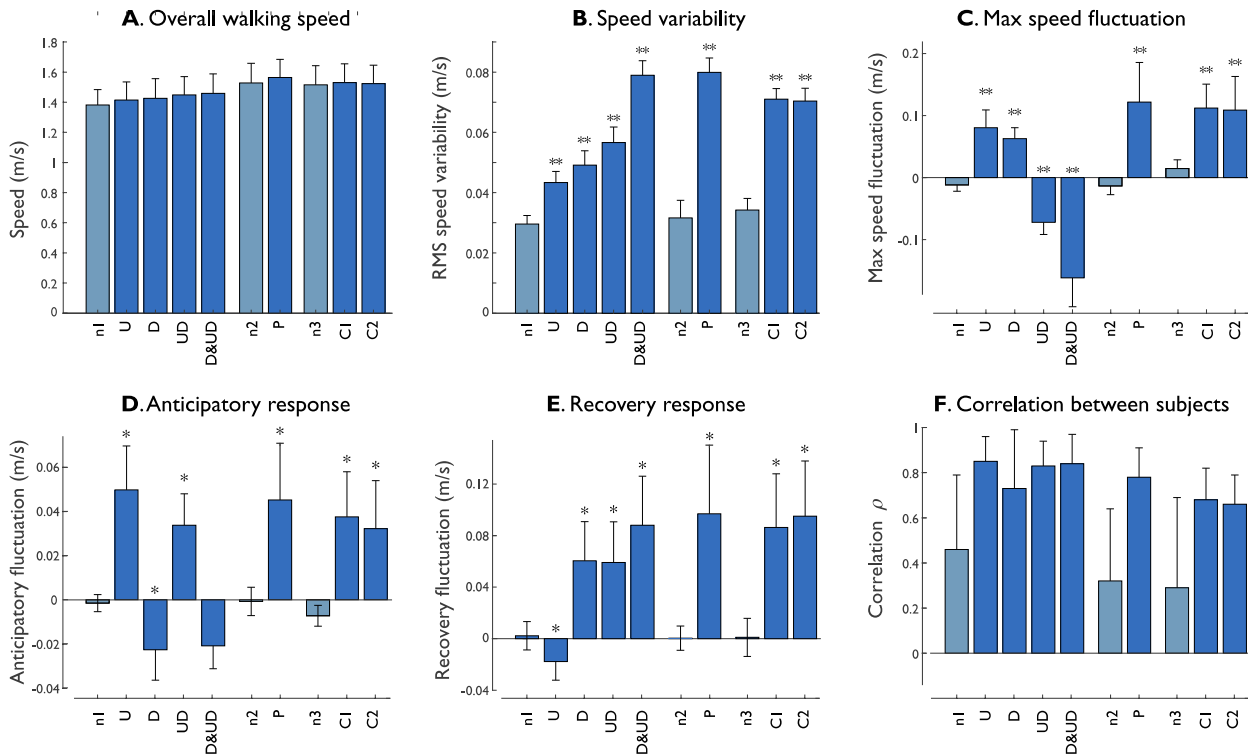

**Figure 6:** Walking statistics for each terrain. (A.) Overall walking speed for three sets of uneven terrain, each preceded by its own level control condition (nominal n1, n2, n3). There were no significant differences in speeds within each set (all P > 0.05). Overall speed was defined as walking distance divided by elapsed time to traverse the terrain, starting and ending at level gait. (B.) Speed variability for each terrain, defined as root-mean-square (RMS) variability of speed as it fluctuated within each trial. (C.) Maximum speed fluctuation for each terrain, defined as the largest observed deviation of speed from nominal. These occurred at terrain-specific step number $i = -4, -1, 2, 1, 3, 17, 9, 11, 16, 16$ respectively for n1, U, D, UND, D&UD, n2, P, n3, C1, C2). (D.) Anticipatory and (E.) Recovery speed fluctuations for each terrain, defined as average deviation of speed (from steady) for two steps immediately preceding or following (respectively) uneven terrain. (F.) Inter-subject correlation for each terrain, between each subject's speed fluctuations and the average pattern across all subjects. Bar graphs show means across subjects; error bars denote s.d. Asterisks denote statistically significant (P < 0.05) differences: * from zero, double asterisks ** from respective control.

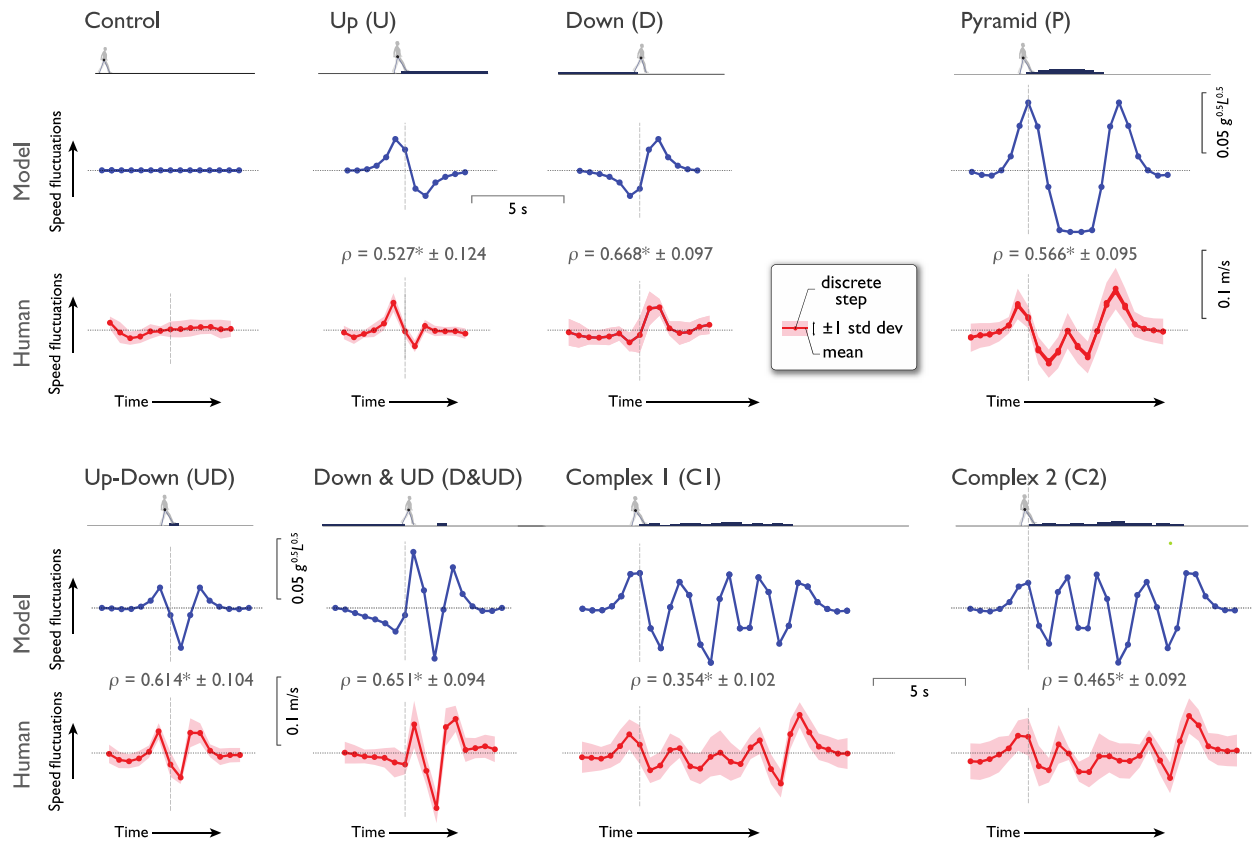

**Figure 7.** Comparison of model and human walking speed fluctuations vs. time, for all terrains. Each terrain (Control, U, D, P, UD, D&UD, C1, C2) is shown with paired Model and Human (top and bottom, respectively; Human $N \geq 11$) trajectories, along with the correlation coefficient $\rho$ between the two (± 95% CI; asterisk * indicates statistical significance, $P < 0.05$). Model predictions are from full-horizon, minimum-energy model (Fig. 3), rendered here in terms of discrete body speed at each step, for comparison with Human average trajectory of body speed (shaded area represents ±1 s.d. from Fig. 5). Each data point corresponds to body speed at a footfall, defined as stride length divided by stride time ending at that footfall. The first uneven step is indicated by vertical dashed line, and overall walking speed is denoted by horizontal solid line. Model trajectories are plotted in dimensionless speed and time, equivalent to units of $g^{0.5}L^{0.5} = 3.13$ m/s and $g^{-0.5}L^{0.5} = 0.32$ s, respectively using gravitational acceleration $g$ and human leg length $L = 1$ m.

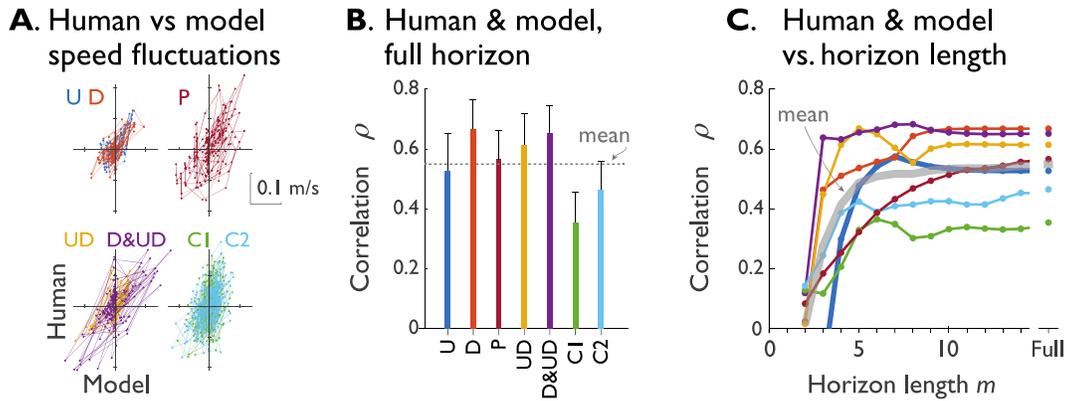

**Figure 8.** Correlation between human and model based on full and finite horizon predictions. (A.) Human vs. model speed fluctuations for each terrain (all subjects, $N \geq 11$), using full-horizon, minimum-energy prediction. The consecutive steps (lines; same data as Fig. 5) by each person on a terrain (filled symbols for individual steps), plotted against model full horizon prediction. Speed fluctuation trajectories have been de-meaned and scaled per subject for visualization purposes; correlation measures are independent of scale). Human speed fluctuations are expected to be positively correlated with model. (B.) Correlation $\rho$ between human speed fluctuations and minimum-energy model prediction, based on full horizon optimization. Terrains are Up (U), Down (D), Pyramid (P), Up-Down (UD), Down-and-Up-Down (D&UD), Complex 1 (C1), and Complex 2 (C2). All correlations are significantly non-zero (P < 0.05; error bars denote 95% CI), as reported in Fig. 7. Also shown is mean correlation across terrains (dotted line, $\rho = 0.55$). (B.) Correlation between human and finite horizon model vs. horizon length, for all terrains (one line per terrain; mean across terrains in gray). A horizon length of about six to eight steps is sufficient to predict human responses on most terrains. The color coding in (B.) also applies to (A.) and (C.).

**Supplementary Information for**

Humans plan for the near future to walk economically on uneven terrain

Osman Darici[1,*], Arthur D. Kuo[1,2]

[1]Faculty of Kinesiology and [2]Biomedical Engineering Program, University of Calgary, Calgary, Alberta, Canada

Osman Darici

Email: osman.darici1@ucalgary.ca

**This PDF file includes:**

    Legends for Datasets S1 to S2 and S3 a MATLAB script to read and plot the human and model speeds.

    Modeling software will be shared in a public repository (under preparation).

**Other supplementary materials for this manuscript include the following:**

    Datasets S1 to S2 and S3 a code script

**Dataset S1 (separate file).** Model Optimal Speed Trajectories for each uneven terrain

**Dataset S2 (separate file).** Human Speed Trajectories for each uneven terrain

**S3 (separate file).** A MATLAB script to read and plot the human and model speeds.